\documentclass[10pt]{iopart}
\usepackage{iopams,bm,cite,color,graphicx,mathptmx,subfigure,url}

\newcommand{\op}[1]{\hat{#1}}
\newcommand{\openone}{\leavevmode\hbox{\small1\normalsize\kern-.33em1}}

\begin{document}

\title{Nonlinear cross-Kerr quasiclassical dynamics}

\author{I~Rigas$^{1,2}$, A~B~Klimov$^{3}$, L~L~S\'anchez-Soto$^{1,2}$ and G~Leuchs$^{1,4}$}

\address{$^{1}$ Max-Planck-Institut f\"ur die Physik des Lichts,
  G\"{u}nther-Scharowsky-Stra{\ss}e 1, Bau 24, 91058 Erlangen,
  Germany}

\address{$^{2}$ Departamento de \'Optica, Facultad de F\'{\i}sica,
  Universidad Complutense, 28040~Madrid, Spain}

\address{$^{3}$ Departamento de F\'{\i}sica, Universidad de
  Guadalajara, 44420~Guadalajara, Jalisco, Mexico}

\address{$^{4}$ Department f\"ur Physik, Universit\"{a}t
  Erlangen-N\"{u}rnberg, Staudtstra{\ss}e 7, 91058 Erlangen, Germany}

\date{\today}

\begin{abstract}
  We study the quasiclassical dynamics of the cross-Kerr effect. In
  this approximation, the typical periodical revivals of the
  decorrelation between the two polarization modes disappear and they
  remain entangled. By mapping the dynamics onto the Poincar\'e space,
  we find simple conditions for polarization squeezing. When
  dissipation is taken into account, the shape of the states in such a
  space is not considerably modified, but their size is reduced.
\end{abstract}

\pacs{03.65.Ta,03.65.Fd,42.50.Dv}

\eqnobysec

\section{Introduction}

The optical Kerr effect refers to the intensity-dependent phase shift
that a light field experiences during its propagation through a
third-order nonlinear medium. This leads to a remarkable non-Gaussian
operation that has sparked considerable interest due to potential
applications in a variety of areas, such as quantum nondemolition
measurements~\cite{Braginsky:1968fk,Unruh:1979uq,Milburn:1983kx,
Imoto:1985vn,Alsing:1988ys,Grangier:1998zr,Sanders:1989ly,Konig:2002uq,Xiao:08},
generation of quantum superpositions~\cite{Milburn:1986ve,Yurke:1986qf,
Tombesi:1987bh,Gantsog:1991cr,Wilson-Gordon:1991bs,Tara:1993nx,
Luis:1995kl,Szabo:1996oq,Chumakov:1999tg,Korolkova:2001ij}, 
quantum teleportation~\cite{Vitali:2000fv,Jian:2009dz,Zhu:2011qa}, 
quantum logic~\cite{Turchette:1995fu,Semiao:2005fu,Matsuda:2007ye,
Azuma:2008qo,You:2011pi,Xia:2011ff}, and single-particle 
detectors~\cite{Kok:2002tw,Munro:2005il,Greentree:2009jl}, to cite
only a few.

Enhanced Kerr nonlinearities have been observed in
electromagnetically-induced transparency~\cite{Schmidt:1996oa,
  Imamoglu:1997sh,Werner:1999xd, Dey:2007wb}, Bose-Einstein
condensates~\cite{Hau:1999xz}, cold atoms~\cite{Kang:2003le} and
Josephson junctions~\cite{Castellanos:2007kx,Mallet:2009uq,
Bergeal:465fk}. Additional arrangements involve the Purcell
effect~\cite{Bermel:2007jb}, Rydberg atoms~\cite{Mohapatra:2008xi},
light-induced Stark shifts~\cite{Brandao:2008mq}, and nanomechanical
resonators~\cite{Babourina-Brooks:2008wm}.

Special mention must be made of the role that this cubic nonlinearity has
played in the generation of squeezed light. The first proposals
employed schemes involving a nonlinear
interferometer~\cite{Ritze:1979mb} or degenerate four-wave
mixing~\cite{Heiman:1976cq,Yuen:1979rq}. But quite soon optical fibers
became the paradigm for that purpose~\cite{Levenson:1985fc,
  Levenson:1985ss,Kitagawa:1986qc,Joneckis:1993bd,Sundar:1996ud,
  Schmitt:1998hs}.  However, due to the typically small
values of the nonlinearity in silica glass~\cite{Boyd:1999kc},
Kerr-based fibers need long propagation distances and high powers,
which bring other unwanted effects~\cite{Shelby:1985sw,Elser:2006fh}.

In this paper, we direct out attention to this limit of high intensities, in
which one could expect a classical description to be pertinent. Under
reasonable assumptions, Maxwell's equations lead to a set of coupled
nonlinear Schr\"{o}dinger equations that has long been a useful tool
for depicting the behavior of optical fields in nonlinear dispersive
media.  It has proved valuable in the description of such diverse
phenomena as pulse compression, dark soliton formation, and
self-focusing of ultrashort pulses~\cite{Agrawal:2007lp}.  Yet there
remain nonclassical features that cannot be explained in this classical
manner. To put it differently, at the most basic level, the
propagation of light in a Kerr medium is accompanied by unavoidable
quantum effects.

The considerations thus far indicate that the regime we wish to
explore can be regarded as a problem at the boundary between classical
and quantum worlds. Probably, the transition between both can be best
scrutinized by exploiting phase-space methods~\cite{Lee:1995wa,
  Schroek:1996ta, Schleich:2001bu}.  This opens up the possibility of
gaining some information about the nonclassical behavior with a
quasiclassical description that employs essentially classical
trajectories, while the correct quantum initial state is taken into
account via, e.g., the Wigner function~\cite{Heller:1976yj,
  Lee:1980bf,Balzer:2003wq}.  Despite some problems with the
interpretation, the Wigner function has enjoyed substantial attention
in various domains of physics~\cite{Hillery:1984ez} and has already
been applied to some nonlinear problems in quantum
optics~\cite{Drobny:1997tp,Banaszek:1997km,Bandilla:2000qw,
Stobinska:2008vz,Corney:2008zr}.

The intensity dependence of the refractive index, which is the
hallmark of the Kerr effect, can manifest itself in two different
ways: as a self-phase modulation and as a cross-phase
modulation. Self-phase modulation refers to the self-induced phase
shift experienced by an optical field during its propagation, whereas
cross-phase modulation refers to the nonlinear phase shift of an
optical field induced by another one having a different wavelength,
direction, or state of polarization.

In this paper we focus on the cross-Kerr effect, for it is especially
germane to attain polarization squeezing, a major goal in our
laboratory~\cite{Sizmann:1999if}.  We capitalize on the quasiclassical
approach to re-analyze the light propagation in this case in a very
concise way: after neglecting higher-order fluctuations, we get an
evolution equation for the Wigner function that can be integrated to 
an analytical form. This allows us to study the dynamics of mode
correlations. Since the resulting state in non-Gaussian, the
application of common entanglement
criteria~\cite{Duan:2000ly,Simon:2000ve} becomes problematic, so we
content ourselves with the study of the purity of the reduced states,
which can be carried out in a closed form.

The two-mode Wigner function is appropriately cast in Poincar\'e space
in terms of the phase-space version of the Stokes parameters, which
affords an intuitive picture. Finally, as the Kerr dynamics is
photon-number preserving, the standard models of
dissipation~\cite{Weiss:1999vn} based in coupling the modes to lossy
reservoirs, seem inadequate.  Instead, we allow for dissipation
through pure dephasing processes which turns out to be exactly
solvable. The resulting evolution reveals that the shape of the Wigner
functions is not considerably modified, but their size is shrunk.

\section{Cross-Kerr quasiclassical evolution}

As heralded in the Introduction, the cross-Kerr configuration
corresponds to a situation in which the refractive index of a beam
(say $a$) is modified by the intensity of a second one (say
$b$). These beams are excited in two orthogonal polarization modes
that, in a quantum description, are characterized by two complex
amplitude operators, denoted $\op{a}$ and $\op{b}$,
respectively. These operators obey the standard bosonic commutation
relations
\begin{equation}
  \label{eq:abcr}
  [ \op{a}, \op{a}^{\dagger} ] =   \op{\openone} = 
  [ \op{b}, \op{b}^{\dagger} ]  \, ,
  \qquad
  [ \op{a}, \op{b} ] = 0 \, ,
\end{equation}
the superscript $\dagger$ standing for the adjoint.

In terms of these annihilation and creation operators, the Hamiltonian
accounting for the cross-Kerr interaction is~\cite{Scully:2001lp}
\begin{equation}
  \label{eq:exactHamiltonian}
  \op{H} = \hbar \chi \; \op{a}^{\dag} \op{a} \op{b}^{\dag} \op{b} \, ,
\end{equation}
where $\chi$ is an effective coupling constant that depends on the
third-order nonlinear susceptibility. For any state described by the
density operator $\op{\varrho}$, the evolution is given by
\begin{equation}
  \label{eq:evol}
  i \hbar \partial_{t} \op{\varrho} = [\op{H}, \op{\varrho}] \, ,
\end{equation}
whose solution can be formally written as
\begin{equation}
  \label{eq:solevol}
  \op{\varrho} (t) = \exp (-i t  \op{H}/\hbar ) 
  \, \op{\varrho} (0) \, \exp ( i t  \op{H}/\hbar ) \, .
\end{equation}
By expanding this equation in the two-mode Fock basis $|n_{a}, n_{b}
\rangle$, the evolution may be, in principle, tracked. Take the
example of an initially pure, two-mode coherent state $| \Psi (0)
\rangle = | \alpha_{0}, \beta_{0} \rangle$, where henceforth the
subscript $0$ indicates the value of the corresponding variable at
$t=0$. The resulting state is
\begin{eqnarray}
  \label{eq:evolexaCohst}
 |\Psi (t) \rangle = \exp (- i t \op{H}/\hbar) | \Psi (0) \rangle
  \nonumber \\ 
  = \exp [ - ( |\alpha_{0}|^2 + |\beta_{0}|^2)/2 ]   
  \sum_{n_{a}, n_{b} =0}^\infty
  \frac{\alpha_{0}^{n_{a}} \beta_{0}^{n_{b}} }{\sqrt{n_{a}!\, n_{b}!}} 
  \, \exp (- i \chi t  n_{a} n_{b} ) |n_{a}, n_{b} \rangle \, .
\end{eqnarray}
The term $\exp (- i \chi t n_{a} n_{b} )$ arises because of the coupling
between the  modes and causes that the state  cannot be factorized
into single-mode states; i.e., it becomes entangled, as we
shall examine in the next section.

It is apparent that equation (\ref{eq:evolexaCohst}) is of practical
use only for few-photon states. Actually, such an exact solution does
not facilitate to extract the classical part of the dynamics in a
manifest form. To that end, we proceed to decompose the mode operators
$\op{a}$ and $\op{b}$ as
\begin{equation}
  \label{eq:abdec}
  \op{a} = \alpha + \delta \op{a} \, , 
  \qquad
  \op{b} = \beta + \delta \op{b} \, , 
\end{equation}
that is,  a sum of classical amplitudes and quantum noise
operators. The average values of the noise operators are assumed
to be much smaller than the corresponding coherent amplitudes ($|
\alpha |^{2} , | \beta |^{2} \gg 1$), so we can restrict the
analysis to first-order terms in $ \delta \op{a} $ and $ \delta
\op{b}$.  If we employ the two-mode Wigner function $ W(\alpha,
\beta)$ and the basic techniques outlined in~\ref{ap:Wigner},
equation~(\ref{eq:solevol}), with this linearization \emph{ansatz},
can be recast as
  \begin{equation}
    \label{eq:Wapprox}
    i \partial_{t} W  =  \chi  |\beta |^{2}
    \left (
      \alpha^{\ast} \frac{\partial W}{\partial \alpha^{\ast}} -
      \alpha \frac{\partial W} {\partial \alpha} 
    \right )  
    +  
    \chi   |\alpha |^{2}
    \left ( 
      \beta^{\ast} \frac{\partial W}{\partial \beta^{\ast}} -
      \beta \frac{\partial W}{\partial \beta}
    \right ) \, .
  \end{equation}
Two comments are in order here. First, we are ignoring quantum
fluctuations, inasmuch  we are disregarding higher-order moments of the
noise operators; this seems a plausible approximation for
highly-excited fields. Second, we underline that the evolution is
specified only by classical trajectories, much in the spirit of the
quasiclassical approximation.

\begin{figure}
  \centering
  \includegraphics[height=5cm]{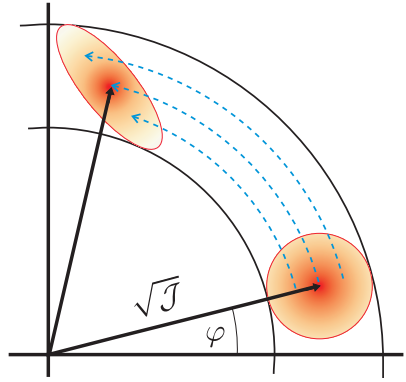}
  \caption{\label{fig:Squeezing} Schematic representation, in the
    phase space of a single mode, of the effect of a Kerr medium. The
    initial state is a coherent state, represented by a circle noise
    determined by the uncertainty relation. It experiences a rotation
    of angle depending on the different amplitudes
    $\sqrt{\mathcal{I}}$. The final result is an elliptical noise
    distribution.}
\end{figure}

To shed light on the physics embodied in~(\ref{eq:Wapprox}),
we resort to action-angle variables $(\mathcal{I}, \varphi)$ for each
mode~\cite{Goldstein:1980fk}. In our context, they can be defined as 
\begin{equation}
  \label{eq:3}
  \alpha = \sqrt{\mathcal{I}_{a}} \, \exp ( i \varphi_{a} ) \, ,
  \qquad 
   \beta = \sqrt{\mathcal{I}_{b}} \, \exp ( i \varphi_{b} ) \, ,
\end{equation}
therefore $\varphi$ is the polar angle in phase space, whereas
$\mathcal{I}$ is related to the mode intensity (see
figure~\ref{fig:Squeezing}). With these variables,
equation~(\ref{eq:Wapprox}) can be rewritten in a simple and elegant
form
\begin{equation}
  \partial_{t} W = \chi \mathcal{I}_{b}
  \frac{\partial W}{\partial \varphi_{a}} + 
  \chi \mathcal{I}_{a}
  \frac{\partial W}{\partial \varphi_{b}} \, .  
  \label{eq:W_p}
\end{equation}
As $\partial/\partial \varphi$ generates rotations in phase space,
equation~(\ref{eq:W_p}) reflects that the amplitudes in each mode
experience different rotations, with angles proportional to the
intensity components of the other
mode~\cite{Heersink:2003oz,Heersink:2005ul}. The result is schematized
in figure~\ref{fig:Squeezing}: roughly speaking, the shaded area
indicates the region in phase space occupied by the state. For an
initial coherent state this area is a circle; the top of the circle
corresponds to higher intensity and therefore is more phase shifted
than the bottom, resulting in an elliptical noise distribution.

Equation~(\ref{eq:W_p}) can be readily solved:
\begin{eqnarray}
  \label{eq:Wigner-T}
  W (\mathcal{I}_{a},\varphi_{a} ; \mathcal{I}_{b} ,\varphi_{b} |t) =  
  W (\mathcal{I}_{a}, \varphi_{a} + \mathcal{I}_{b} \chi t;
  \mathcal{I}_{b} , \varphi_{b}  + \mathcal{I}_{a} \chi t|0) \, .
\end{eqnarray}
If again we assume initially the two-mode coherent state $|\alpha_{0},
\beta_{0} \rangle$ ($\alpha_{0} = \sqrt{\mathcal{I}_{0a}} \,
e^{i\varphi_{0a}}$, $\beta_{0} = \sqrt{\mathcal{I}_{0b}} \, e^{i
  \varphi_{0b}}$) and using (\ref{eq:Wcoh}),
equation~(\ref{eq:Wigner-T}) reduces to
\begin{eqnarray}
  \displaystyle 
  W (\mathcal{I}_{a}, \varphi_{a} ; \mathcal{I}_{b},\varphi_{b} | \tau) 
  &  = & 
  \frac{4}{\pi^2}
  \exp  \left [ -2 | \sqrt{\mathcal{I}_{a}} e^{i(\varphi_{a}  + 2
      \mathcal{I}_{b} \tau )} 
    - \sqrt{\mathcal{I}_{0a}} e^{i\varphi_{0a}} |^{2} \right ] 
  \nonumber \\
  & \times &
  \exp \left [ - 2 | \sqrt{\mathcal{I}_{b}} e^{i (\varphi_{b} 
      + 2 \mathcal{I}_{a} \tau )}  - \sqrt{\mathcal{I}_{0b}}
    e^{i\varphi_{0b}}|^{2} \right ] \,  ,
  \label{Wt_0}
\end{eqnarray}
where we have introduced the dimensionless variable $\tau = \chi t
/2$. Observe that at $ \tau = 0$ the Wigner function is made of two
independent Gaussians, while as time goes by the induced mode
correlations lead to a non-Gaussian state.

\section{Mode correlation dynamics}

Two-mode Gaussian states constitute the simplest example of a
continuous-variable bipartite system, the workhorses of quantum
information.  Accordingly, the theoretical aspects of these states
have been extensively worked out and a variety of quantitative
characterizations are available for them~\cite{Braunstein:2005fk,
Paris:2005uq,Wanga:2007ys,Adesso:2007zr,Weedbrook:2012kx}.

The unique feature of these Gaussian states is that they are fully
specified (up to local displacements) by the covariance matrix
$\bm{\gamma}$, with elements $\gamma_{ij} = \Tr [ \op{\varrho} \{
\op{R}_{i}, \op{R}_{j} \} /2 ]$, where $\{ , \}$ denotes the
anticommutator and $\op{\mathbf{R}} = (\op{x}_{a}, \op{p}_{a},
\op{x}_{b}, \op{p}_{b} )$ is the vector of phase-space operators.
 This covariance matrix can be jotted down as
\begin{equation}
  \label{eq:CM1}
  \bm{\gamma} = 
 \left (
\begin{array}{cc}
\mathbf{A} & \mathbf{C} \\
\mathbf{C}^{t} & \mathbf{B} 
\end{array}
\right ) \, .
\end{equation}
Here, $\mathbf{A}$ and $\mathbf{B}$ are the covariance matrices
associated to the reduced state of the modes $a$ and $b$, while
$\mathbf{C}$ describes the correlation between these modes.  The
symplectic eigenvalues of $\bm{\gamma}$ are
\begin{equation}
  \label{eq:symeigp}
  \nu_{\pm}^{2} =  \frac{1}{2} \left [ \Delta \pm \sqrt{\Delta^{2} - 4
      \det \bm{\gamma}} \right ] \, ,
\end{equation}
with $\Delta = \det \mathbf{A} + \det \mathbf{B} + 2 \det
\mathbf{C}$. These symplectic eigenvalues encode all the essential
information and provide powerful, simple ways to express fundamental
properties. For example, a Gaussian state is entangled if and only if
\begin{equation}
  \label{eq:symeig}
  \tilde{\nu}_{-} < 1/2 \, ,
\end{equation}
where the smallest symplectic eigenvalue $\tilde{\nu}_{-}$ of the
covariance matrix corresponding to the partially transposed state is
obtained from $\nu_{-}$ by replacing $\det \mathbf{C}$ with $- \det
\mathbf{C}$; i.e., by time reversal in the second system and thus a
flip of its canonical momentum.

\begin{figure}
  \centering
  \includegraphics[height=6.5cm]{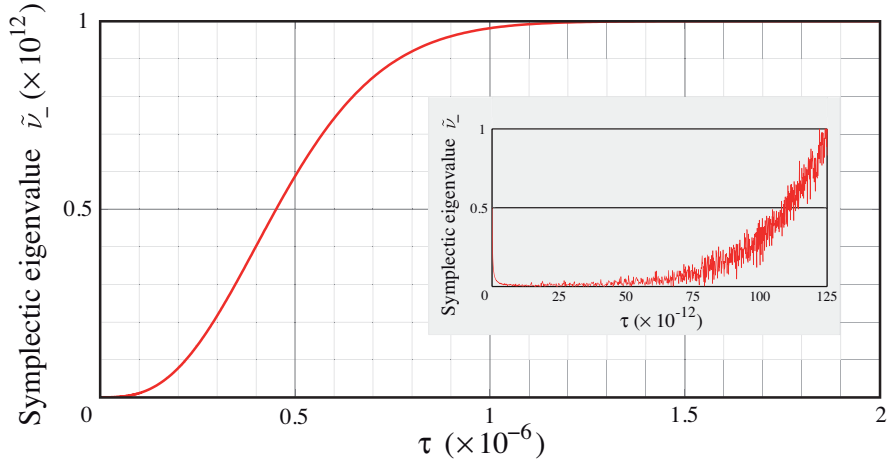}
  \caption{\label{fig:symeig} Time evolution of the symplectic
    eigenvalue $\tilde{\nu}_{-}$ of the state (\ref{Wt_0}), as a
    measure for the entanglement between the two modes. We have taken
    both modes with the same intensity $\mathcal{I}_{0a} =
    \mathcal{I}_{0b} = 10^{6}$. Entanglement is proven for
    $\tilde{\nu}_{-} < 0.5$, a region which can be observed only in
    the inset.}
\end{figure}

In figure~\ref{fig:symeig} we have plotted the time evolution of
$\tilde{\nu}_{-}$ for the state (\ref{Wt_0}), showing a rapid increase
(in the inset, we observe a fluctuating behavior that is smeared out
in a larger scale).  The main caveat with this approach is that, as
mentioned before, our state rapidly becomes non-Gaussian, and
criterion (\ref{eq:symeig}) gives then only a sufficient
condition. Consequently, we can certify entanglement just in the
short-time window displayed in the inset.  This actually holds for any
available criterion~\cite{Duan:2000ly,Simon:2000ve}: if the state is
entangled, a given test may or not detect its entanglement; in turn,
if a particular test does not detect entanglement, we can not conclude
separability of the state.

Genuine non-Gaussian entanglement can only be revealed by measures
involving higher-order moments. In this vein, Shchukin and
Vogel~\cite{Shchukin:2005bh} (see
also~\cite{Agarwal:2005dq,Hillery:2006nx}) have introduced a general
hierarchy of necessary and sufficient conditions for any state to be
entangled. Nevertheless, the application of this technique to our
problem turns out to be very arduous for it involves checking
non-trivial inequalities, which can be performed only
numerically. Moreover, the method involves the determination of
moments that are extremely oscillatory and noisy~\cite{Gomes:2009qf}.

In view of these difficulties, we content ourselves with assessing the
purity of the reduced state of both modes.  This is related to the
linear entropy and intimately connected to the intermodal
correlations~\cite{Mazzola:2010ly}. These local purities are 
\begin{equation}
  \label{eq:defpur}
  P_{a} (\tau) = \Tr_{a} [ \op{\varrho}_{a}^{2} (\tau) ] \, , 
  \qquad
  P_{b} (\tau) = \Tr_{b} [ \op{\varrho}_{b}^{2} (\tau) ] \, ,
\end{equation}
$\op{\varrho}_{a} (\tau) = \Tr_{b} [ \op{\varrho} (\tau) ]$ and
$\op{\varrho}_{b} (\tau) = \Tr_{a} [ \op{\varrho} (\tau) ]$ being the
reduced density matrices of modes $a$ and $b$, respectively.  If we
employ now the two-mode Wigner function  (\ref{Wt_0}),
the purity, say $P_{a} (\tau)$, can be written  as
\begin{equation}
  P_{a} (\tau )  = \frac{\pi}{8} \int_{-\pi}^{\pi} d\varphi_{a}
  \int_{0}^{\infty} d\mathcal{I}_{a}  
  \left [ \int_{-\pi}^{\pi} d\varphi_{b} \int_{0}^{\infty} d\mathcal{I}_{b} 
    W(\mathcal{I}_{a}, \varphi_{a} ; \mathcal{I}_{b}, \varphi_{b}| \tau) \right ]^{2} \, .
\end{equation}

For a bipartite system,  both purities in (\ref{eq:defpur})  coincide
for pure states~\cite{De-Pasquale:2010ys,De-Pasquale:2012zr}. In
general, these quantities are different for mixed states.  In our
case,  after a long but otherwise straightforward calculation (which, for
completeness, is sketched in \ref{ap:purity-reduc-dens}),  $P_{a} (\tau)$
can be displayed as
\begin{equation}
  P_{a} ( \tau )  =  \exp (-4\mathcal{I}_{0b} - 2\mathcal{I}_{0a} )
  \sum_{n=-\infty}^{\infty} 
  \frac{I_{n} ( 2\mathcal{I}_{0a} )}{1 +\tau^{2} n^{2}} 
  \exp \left ( \frac{4 \mathcal{I}_{0b}}{1 + \tau^{2} n^{2}} \right )
  = P_{b} (\tau) \, ,  
  \label{Pex}
\end{equation}
where $I_{n} (z)$ are the modified Bessel functions of first kind and
the last equality has been carefully checked by numerical experiments.
This surprising symmetry can be ascribed to the way in which
the modes enter the Kerr Hamiltonian~(\ref{eq:exactHamiltonian}). 
Accordingly, we drop the mode subscripts in the purities.

As we are dealing with highly-excited fields 
($\mathcal{I}_{0a} \gg 1$), we can make use of the asymptotic
expansion~\cite{Abramowitz:1984}
\begin{equation}
  \label{eq:BesselI}
  I_{n} (z) \sim  \frac{e^{z}}{\sqrt{2\pi z}} e^{-n^{2}/2z} \, , 
\qquad 
|z| \gg 1 \, .
\end{equation}
In addition, as $\tau \ll 1$ and the functions in (\ref{Pex}) do not
oscillate, we can replace the summation by an integral; the final
result being
\begin{equation}
  \label{eq:sumbyint}
  P ( \tau ) = \frac{1}{\sqrt{4\pi \mathcal{I}_{0a}}}
  \int_{-\infty}^{\infty} dx 
  \frac{1}{1+ \tau^{2} x^{2}} 
  \exp \left ( -\frac{4 \mathcal{I}_{0b} \tau^{2}x^{2}}
    {1 + \tau^{2} x^{2}} - \frac{x^{2}}{4 \mathcal{I}_{0a}} \right ) \, .
\end{equation}

\begin{figure}
  \centering
  \includegraphics[height=6.5cm]{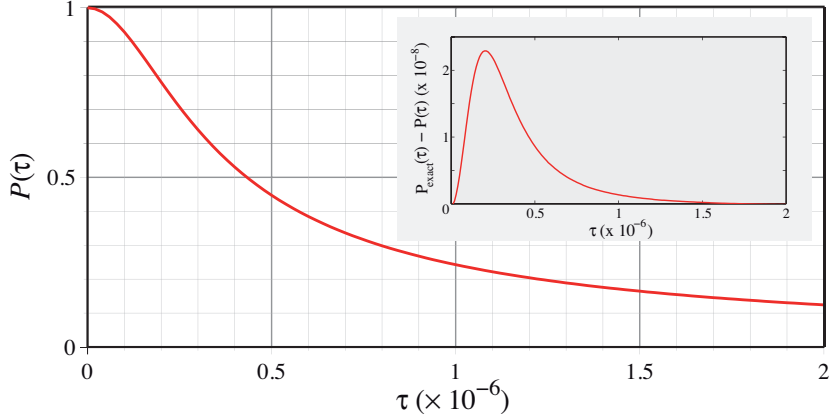}
  \caption{\label{fig:purity} Evolution of the purity $P(\tau)$ as a
    measure for the correlations between the two modes for the same
    conditions as in figure~2.}
\end{figure}

In figure~\ref{fig:purity} we plot the time evolution of this $P
(\tau)$ in the same scale as in figure~\ref{fig:symeig}. At $\tau = 0$
the reduced purity is unity, in agreement with the fact that initially
the state consists of two uncorrelated Gaussians. As time evolves, the
purity smoothly decreases (much in a similar way as the symplectic
eigenvalue $\tilde{\nu}_{-}$ decreases), which indicates the presence
of mode correlations. This is supported by the following asymptotic
estimate of (\ref{eq:sumbyint})
\begin{equation}
  \label{eq:lim}
  P (  \tau )  \simeq  \frac{1}
  {\sqrt{1 + 16 \mathcal{I}_{0a} \mathcal{I}_{0b} \tau^{2}}} \, .
\end{equation}
valid for $\mathcal{I}_{0b} \tau \lessapprox 1$.  It is clear that this form
of $P(\tau)$ is invariant under mode permutations.  Finally, $P
(\tau)$  tends to its stationary value.

One might wonder how quantum fluctuations, neglected thus far, could
modify this quasiclassical picture. For the particular case of initial
coherent states we are treating here, we can analytically compute the
purity for the exact quantum solution. Indeed, from
(\ref{eq:evolexaCohst}) we have
\begin{equation}
  \fl 
  \op{\varrho}_{a} (t) 
  = \exp [-  | \alpha_{0} |^2 - | \beta_{0 }|^2 ] \sum_{n_{a},
    n_{b}=0}^\infty 
  \frac{\alpha_{0}^{n_{a}} \alpha_{0}^{\ast n_{b}}}
  {\sqrt{n_{a}! \,  n_{b}!}} \, 
  \exp \left [ e^{2i\tau (n_{a} - n_{b})} |\beta_{0}|^2 \right] 
  | n_{a} \rangle \langle n_{b}| \, , 
\end{equation}
wherefrom one easily derive the exact expression for the purity:
\begin{equation}
  \label{eq:ExactReducedPurity}
  P_{\mathrm{exact}} ( \tau ) = \exp ( - 2 \mathcal{I}_{0a} - 2 \mathcal{I}_{0b} )
  \sum_{n = -\infty}^\infty
  I_{n} (2 \mathcal{I}_{0b} ) \,
  \exp[ 2 \mathcal{I}_{0a} \cos (2 n \tau ) ] \, . 
\end{equation}
Using the properties of the Bessel functions, we redraft this as
\begin{equation}
  \label{eq:ExactReducedPurity2}
  P_{\mathrm{exact}} ( \tau ) = \exp ( - 2 \mathcal{I}_{0a} - 2 \mathcal{I}_{0b} )
  \sum_{m, n = -\infty}^\infty
    I_{m} (2 \mathcal{I}_{0a} )  I_{n} (2 \mathcal{I}_{0b} ) \, 
  \exp (2 i m   n \tau )  \, , 
\end{equation}
which explicitly exhibits the aforementioned symmetry. In fact, taking into account
(\ref{eq:BesselI}), $ P_{\mathrm{exact}} ( \tau ) $ appears as a
bidimensional Jacobi theta function~\cite{Abramowitz:1984},
which is periodic. However, in the time scales we are considering
here, such a periodicity is unnoticeable and we can replace again the
sum by an integral, getting precisely equation~(\ref{eq:lim}). 

In the inset of figure~\ref{fig:purity} we have plotted the difference
between the exact solution (\ref{eq:ExactReducedPurity2}) and the
quasiclassical one (\ref{eq:sumbyint}). As we can see, both solutions
coincide for any practical purpose. This means that the correlations
examined before are of quantum nature, but higher-order correlations
play no relevant role here.

\section{Polarization squeezing}

Since the polarization modes $a$ and $b$ have the same frequency and
are orthogonal, their superposition results in a general elliptical
polarization.  This means that one needs only three independent
quantities: the amplitudes of each mode and the relative phase between
them. To describe this at the quantum level, it is advantageous to use
the Stokes operators~\cite{Luis:2000ys}
\begin{equation}
  \label{gensu2}
  \op{S}_x  = 
  \op{a}^\dagger \op{b} +
  \op{b}^\dagger \op{a}  \, ,
  \qquad
  \op{S}_y  =  i
  (\op{a} \op{b}^{\dagger} -
  \op{a}^\dagger \op{b} ) \, , 
  \qquad
  \op{S}_z =   
  \op{a}^\dagger \op{a} -
  \op{b}^\dagger \op{b}   \, ,
\end{equation}
complemented with the total number $\op{N} = \op{a}^\dagger \op{a} +
\op{b}^\dagger \op{b}$.  On account of  (\ref{eq:abcr}), the operators
(\ref{gensu2}) satisfy the commutation relations of an angular momentum
\begin{equation}
  \label{crsu2}
  [ \op{S}_{k}, \op{S}_{\ell}] = 2 i \varepsilon_{k\ell m} \,
  \op{S}_{m}  \, , \qquad [\op{N}, \op{S}_{k} ] = 0 \, ,
\end{equation}
where the Latin indices run over $\{ x, y, z \}$ and
$\varepsilon_{k\ell m}$ is the Levi-Civita fully antisymmetric
tensor. This noncommutability precludes the simultaneous exact
measurement of the physical quantities they represent and leads
immediately to the Heisenberg inequalities~\cite{Sehat:2005wd,
  Marquardt:2007bh,Bjork:2010rt,Muller:2012mk}
\begin{equation}
  \label{eq:polsquez1}
  \Delta^{2} \op{S}_{k} \,  \Delta^{2} \op{S}_{\ell}  \ge 
  \epsilon_{k\ell m} \, | \langle \op{S}_{m} \rangle  |^{2} \, ,
\end{equation}
where $\Delta^{2} \op{A} = \langle \op{A}^{2}\rangle - \langle \op{A}
\rangle^{2}$ indicates the variance.  It is always possible to find pairs of
maximally conjugate operators for this uncertainty relation. This is
equivalent to establishing a basis in which only one of the operators
(\ref{gensu2}) has a nonzero expectation value, say
\mbox{$\langle\op{S}_{k} \rangle = \langle \op{S}_{\ell} \rangle=0$}
and $\langle\op{S}_{m} \rangle \neq 0$.  The only nontrivial
Heisenberg inequality reads thus
\begin{equation}
  \Delta^2 \op{S}_{k} \, \Delta^2 \op{S}_{\ell}  \geq 
  |\langle\op{S}_m \rangle|^2 \, .
\end{equation}

Polarization squeezing can then be sensibly defined by the
condition~\cite{Chirkin:1993dz,Korolkova:2002fu,Luis:2006ye,Mahler:2010fk,Ma:2011ve}
\begin{equation}
  \Delta^2 \op{S}_{k} < |\langle\op{S}_{m} \rangle| 
  < \Delta^2 \op{S}_{\ell} \, .
  \label{eq:polsq1}
\end{equation}
Note that squeezed states according to (\ref{eq:polsq1}) are not, in
general, minimum uncertainty states.

The choice of the conjugate operators $\{\op{S}_{k},\op{S}_{\ell} \}$
is by no means unique: there exists an infinite set $\{
\op{S}_\perp(\vartheta), \op{S}_\perp(\vartheta+\pi/2) \}$ that are
perpendicular to the classical excitation $\langle \op{S}_{m}
\rangle$, for which $\langle\op{S}_\perp(\vartheta) \rangle =0$ for
all $\vartheta$. All these pairs exist in the $S_{k}$--$S_{\ell}$
plane, which is called the \textit{dark plane}. A generic
$\op{S}_{\perp} (\vartheta)$ can be written as
\begin{equation}
  \label{eq:Sdark}
  \op{S}_{\perp} (\vartheta) = \op{S}_{k} \, \cos \vartheta + 
 \op{S}_{\ell} \sin \vartheta \, ,
\end{equation}
$\vartheta$ being an angle defined relative to $\op{S}_{k}$.
Condition (\ref{eq:polsq1}) is then equivalent to
\begin{equation}
  \Delta^2 \op{S}_\perp (\vartheta_{\mathrm{sq}}) < 
  \textstyle | \langle \op{N} \rangle| < 
  \Delta^2 \op{S}_\perp ( \vartheta_{\mathrm{sq}} + \pi/2 ),
  \label{eq:polsq2}
\end{equation}
where $\op{S}_\perp(\theta_{\mathrm{sq}} )$ is the maximally squeezed
operator and $\op{S}_\perp( \theta_{\mathrm{sq}} + \pi/2 )$ the
antisqueezed one.

In many experiments both modes have the same amplitude but are phase
shifted by $\pi / 2$: $\langle \op{a} \rangle = i \langle
\op{b}\rangle$.  This light is circularly polarized and fulfills
$\langle \op{S}_{x} \rangle = \langle \op{S}_{z} \rangle = 0$,
$\langle \op{S}_{y} \rangle \neq 0$, so (\ref{eq:polsq2}) directly
applies.

The time evolution of the variables involved in those definitions can
be evaluated using the Wigner-distribution approach:
\begin{equation}
  \langle \op{S}_{\perp} (\vartheta, \tau ) \rangle =\pi^{2} \mathrm{Re} 
  \left [ e^{i\vartheta} \int d^{2}\alpha d^{2}\beta \, 
    W_{\op{S}_\perp (\vartheta)}(\alpha ,\beta) \, W (\alpha ,\beta | \tau)
  \right] \, .
\end{equation}
Here, $W_{\op{S}_\perp ( \vartheta)}(\alpha ,\beta)$ refers to the
phase-space function corresponding to the operator $\op{S}_\perp
(\vartheta)$ (commonly called its symbol).  From (\ref{gensu2}) and
(\ref{eq:Sdark}) it is clear that the symbol of $\op{S}_\perp
(\vartheta)$ can be directly constructed in terms of the symbols of
the basic mode amplitudes $\op{a}$ and $\op{b}$, which, from
\ref{ap:Wigner}, we know are given by $W_{\op{a}} (\alpha ) =
\alpha/\pi$ and $W_{\op{b}} (\beta ) = \beta/\pi$. Therefore, we get
\begin{equation}
  \label{eq:SyOfT}
  \langle \op{S}_{y} ( \tau ) \rangle =
  \frac{\mathcal{I}_{0}}{( 1+ \tau^{2} )^{2}}
  \exp \left ( - \frac{2 \mathcal{I}_{0} \tau^{2}}
    {1 + \tau^{2}} \right )  \, ,
  \qquad
  \langle \op{S}_{x}( \tau ) \rangle = 
  \langle \op{S}_{z}( \tau ) \rangle = 0\, ,
\end{equation} 
where $\mathcal{I}_{0} = \Tr [ \op{\varrho} (0) \op{N} ] $ is the
initial average number of photons of the state.  The second-order
moments are calculated much in the same way; the final result being
\begin{eqnarray}
  \label{eq:DeltaSTheta}
  \fl
  \Delta^{2} \op{S}_{\perp} ( \vartheta, \tau)  & =  &
  \mathcal{I}_{0} [ 1 + (\mathcal{I}_{0}/2) \sin^2 (\vartheta /2 ) ]  
  -\sin^2\vartheta \frac{2 \mathcal{I}_{0}^{2}}{(1+4\tau^2)^3}
  \exp\left(  -\frac{8 \mathcal{I}_{0} \tau^2}{1 + 4\tau^2}\right) \nonumber\\
  \fl 
  & - &\displaystyle 
  \sin (2\vartheta)\frac{2 \mathcal{I}_{0} \tau}{(1 + \tau^2)^3} 
  \left( 1+ \frac{\mathcal{I}_{0} }{1 + \tau^2} \right)
  \exp \left(
    -\frac{2 \mathcal{I}_{0 }\tau^2}{1 + \tau^2}\right) \, .
\end{eqnarray}
A major advantage of this formalism is that we can specify the time
evolution of polarization squeezing.  In particular, for sufficiently
short times $\tau \ll 1 $, we can expand equations~(\ref{eq:SyOfT})
and (\ref{eq:DeltaSTheta}) up to second order, so that
\begin{eqnarray}
  \label{eq:sdap}
  \fl
  \langle \op{S}_y ( \tau ) \rangle \simeq  
  \mathcal{I}_{0} (1 - \mathcal{I}_{0} \tau^2 ) \, , 
  \quad
  \Delta^{2} \op{S}_{\perp} (\vartheta,  \tau )  \simeq  
 \mathcal{I}_{0} [ 1 + 4 \mathcal{I}_{0}^2 \, 
  \sin^2 (\vartheta) \, \tau^2] - 2 \mathcal{I}_{0}^2  \sin(2\vartheta)  \tau \, ,
\end{eqnarray}
so that the optimal squeezing angle is roughly given by
\begin{equation}
  \label{eq:OptSqzAngle}
  \vartheta_{\mathrm{sq}} \simeq   \frac{1}{2}
  \,\mathrm{arccot} ( \mathcal{I}_{0}  \tau ) \, ,
\end{equation}
i.e., it starts at $\vartheta_{\mathrm{sq}} = \pi/4$ and slowly moves
towards $0$ as $\tau$ goes by.

\section{Mapping the dynamics on the sphere}

It is possible to turn the action the Stokes operators discussed in
the previous section into a very simple phase-space picture. To this
end we introduce the parametrization~\cite{Klimov:2006fd}
\begin{equation}
  \label{eq:ppar}
  \alpha = \sqrt{\mathcal{I}} \, e^{i \varphi_{a}}  \, \cos (\theta/2) \,  ,
  \qquad
  \beta = \sqrt{\mathcal{I}} \, e^{i \varphi_{a}} e^{- i \phi}  \, \sin (\theta /2) \, ,
\end{equation}
where $\varphi_{a}$ appears now as a global phase and the pertinent
relative phase is $\phi = \varphi_{a} - \varphi_{b}$. The radial
variable
\begin{equation}
  \mathcal{I} = \mathcal{I}_{a} + \mathcal{I}_{b}
\end{equation}
represents the total intensity. The parameters $\theta$ and $\phi$ can
be interpreted as the polar and azimuthal angles, respectively, on the
Poincar\'e sphere: $\theta$ describes the relative amount of intensity
carried by each mode and $\phi$ is the relative phase between them. In
term of these new variables, equation~(\ref{eq:W_p}) becomes
\begin{equation}
  \partial_{t} W =  
  \chi  \mathcal{I}_{b}    \frac{\partial W}{\partial \varphi_{a}} +
  \chi  ( \mathcal{I}_{b} - \mathcal{I}_{a} ) \frac{\partial W}{\partial \phi} \, .
\end{equation}
In (\ref{eq:ppar}), $\varphi_{a}$ appears as an irrelevant global
phase over which we can integrate without loosing relevant
information; the result is
\begin{equation}
  \partial_{t} W (\mathcal{I} , \theta, \phi) = -\chi \mathcal{I} \cos \theta 
  \frac{\partial W (\mathcal{I}, \theta , \phi)}{\partial \phi},  
  \label{Wav}
\end{equation}
whose solution in terms of the adimensional variable $\tau$ reads
\begin{equation}
  \label{eq:Wdlh}
  W (\mathcal{I} , \theta , \phi | \tau ) = 
  W (\mathcal{I},  \theta, \phi - 2 \tau \mathcal{I}  \cos\theta  | 0 ) \, .
\end{equation}

The three numbers $(\mathcal{I}, \theta, \phi )$ are the spherical
coordinates in the Poincar\'e space:
\begin{equation}
  \label{eq:Sdefs}
  S_{x} = \mathcal{I} \sin\theta\cos\phi \, , 
  \qquad 
  S_{y} = \mathcal{I}\sin\theta\sin\phi \,,
  \qquad
  S_{z} = \mathcal{I} \cos\theta\,  .
\end{equation} 
In terms of the Cartesian counterpart equation~(\ref{eq:Wdlh}) can be
compactly expressed as
\begin{equation}
  \label{eq:Wigner-S|t}
  W (S_{x}, S_{y}, S_{z} | \tau) = \frac{8}{\pi} \, \exp(-2\mathcal{I} - 2\mathcal{I}_{0} ) \,
  I_0 \left ( 2 \sqrt{\sigma(\theta,\phi,\tau)} \right )  \, ,
\end{equation}
where
\begin{equation}
  \label{eq:sigma} 
  \fl
  \sigma(\theta,\phi,\tau)  =  2 
  \left [ \mathcal{I}\,\mathcal{I}_{0} + S_z S_{0z} +  
    \cos( 2 S_z\tau)(S_x S_{0x} + S_y S_{0y})  + 
    \sin( 2 S_z \tau) (S_y S_{0x} - S_x S_{0y}) \right ] \, .
\end{equation}
For the aforementioned case of circularly polarized light, with $S_{0x} = S_{0z} = 0,
S_{0y} =\mathcal{I}_{0}$, this reduces to
\begin{equation}
  \label{eq:Sapprox}
  \sigma(\theta,\phi,\tau) = 2 \mathcal{I}_{0}
  \left [  
    \mathcal{I} +  S_y \,\cos(2S_z \tau) -  S_x\,\sin(2S_z\tau)
  \right ] \,.
\end{equation}
In the $x$-$p$ quadrature phase space, the usual way of representing
states is by an uncertainty region which is just a contour of the
Wigner function $W(x,p)$ for that state. Much in the same way, for
each fixed time, the equation $W (S_{x}, S_{y}, S_{z} | \tau ) = $
constant defines an isocontour surface in the Poincar\'s space of axes
($S_{x}$, $S_{y}$, $S_{z}$), which gives complete information about
the fluctuations of the state. In the supplementary material of this
paper, we include a movie portraying the time evolution of the Wigner
function (\ref{eq:Wigner-S|t}) for the particular instance in
(\ref{eq:Sapprox}).  As it can be appreciated, the state gets
elongated along the direction of maximal squeezing. In figure~4 we
present three snapshots of the movie, corresponding to different
times.
 
\begin{figure}
  \centering
  \includegraphics[width=0.95\columnwidth]{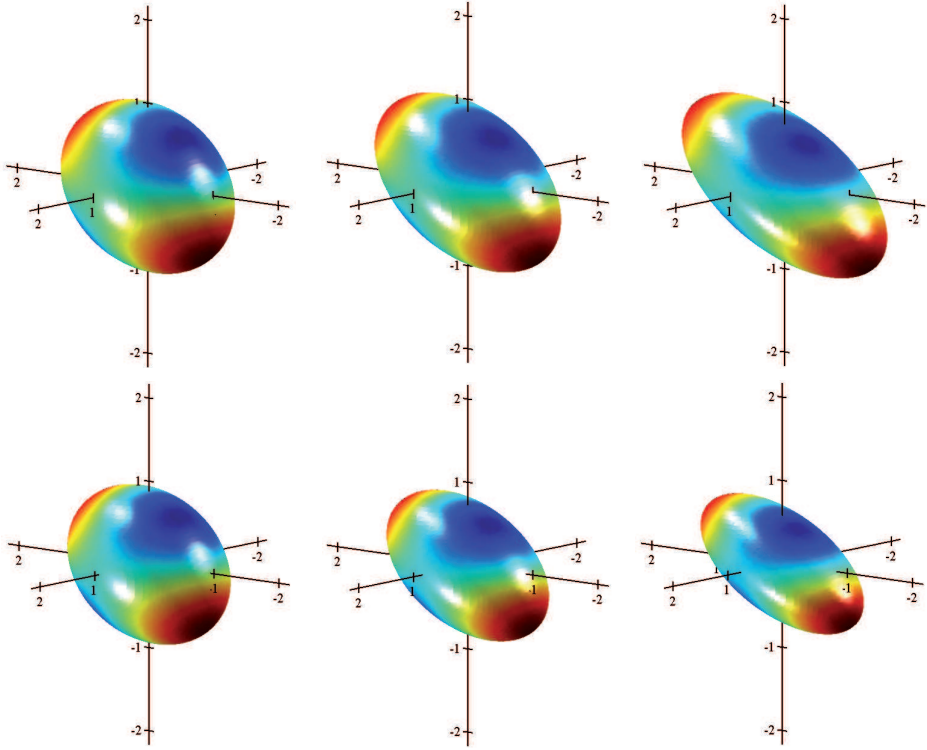}
  \caption{Isocontour surfaces of the level $10^{-4}$ (from the
    maximum) of the Wigner function $W (S_{x}, S_{y}, S_{z} | \tau ) = $
    constant at times $\tau = 1.5\times 10^{-7}, 3.0 \times 10^{-7}$
    and $4.5 \times 10^{-7}$ (from left to right), without dephasing
    (top) and with a dephasing of $\gamma = 0.5 \chi$ (bottom). The
    orthogonal axis are $S_{x}$, $S_{y}$ and $S_{z}$, the box is
    centered at $S_x = S_z = 0, S_y = 10^6$ and the axis ticks are
    measured in unit of the (spherical) isocontour at $\tau = 0$,
    which corresponds to the shot-noise limit.}
\end{figure}

\section{Dissipative effects}

As light propages through the Kerr medium, it experiences a
decorrelation of the relative phase between both basic polarization
modes.  A sensible approach to deal with this decorrelation is through
the notion of decoherence, by which we loosely understand the
appearance of irreversible and uncontrollable quantum correlations
when a system interacts with its environment~\cite{Zurek:2003fk}.

Usually, decoherence is accompanied by dissipation, i.e., a net
exchange of energy with the environment. However, giving the nature of
the Kerr nonlinearity, we are interested in the case of pure
decoherence (also known as dephasing), for which the processes of
energy dissipation are negligible.  Models in which the number of
photons do not change, while the coherences are strongly decaying, are
at hand~\cite{Shao:1996uq,Mozyrsky:1998kx,Gardiner2004,Breuer2002,
Klimov:2008ys}. Surprisingly enough, however, they have not been
applied in the context of the phase-number preserving Kerr
dynamics. In consequence, we model such a dephasing by the master
equation
\begin{equation}
  \partial_{t} \op{\varrho} = -i [\op{H},\op{\varrho}] + 
  \gamma_{a} \mathcal{L}_{\op{a}} [ \op{\varrho} ] + 
  \gamma_{b} \mathcal{L}_{\op{b}} [\op{\varrho} ] \, ,
\end{equation}
where $\mathcal{L}_{\op{a}} [ \op{\varrho} ]$ is the Linblad
superoperator
\begin{equation}
  \label{eq:Lindeph}
  \mathcal{L}_{\op{a}} [ \op{\varrho} ] = 2 
  \op{a}^{\dag} \op{a} \, \op{\varrho} \, \op{a}^{\dag}\op{a} -
  ( \op{a}^{\dag} \op{a} )^{2} \, \op{\varrho} - 
  \op{\varrho} \, ( \op{a}^{\dag} \op{a} )^{2} \, ,
\end{equation}
with $\gamma_{a}$ the dephasing constant. A similar expression holds
for mode $b$.  The equation for the Wigner function (\ref{eq:W_p}) is
modified now to
\begin{equation}
  \partial_{t}W  =  \chi 
  \mathcal{I}_{b} 
  \frac{\partial W}{\partial \varphi_{a}} + 
  \chi 
    \mathcal{I}_{a}
  \frac{\partial W}{\partial \varphi_{b}}  
  + \frac{\gamma_{a}}{4} 
  \frac{\partial^{2}W}{\partial \varphi_{a}^{2}} + 
  \frac{\gamma_{b}}{4} \frac{\partial^{2}W}{\partial \varphi_{b}^{2}} \, .
\end{equation}
Using again the variables (\ref{eq:Sdefs}) and integrating over the
irrelevant overall phase $\varphi_a$, this equation turns out to be
\begin{equation}
  \partial_{t} W (\mathcal{I}, \theta , \phi ) = -\chi \mathcal{I} \cos \theta 
  \frac{\partial W (\mathcal{I}, \theta , \phi)}{\partial \phi} + 
  \frac{\gamma}{4}
  \frac{\partial^{2} W (\mathcal{I}, \theta , \phi )}{\partial \phi^{2}},
\end{equation}
with $\gamma = \gamma_a + \gamma_b$. Its general solution can be
represented by
\begin{equation}
  W (\mathcal{I}, \theta , \phi | t)  = \frac{1}{2 \pi} \int  d\phi^{\prime} 
  \Theta (\phi - \phi^{\prime}-  \chi t  \mathcal{I} \cos \theta| t\gamma/4) \,
  W (\mathcal{I}, \theta,  \phi^{\prime} | 0) \, ,
  \label{eq:DiffSolution}
\end{equation}
with $\Theta (\phi |t) = \sum_{k} \exp (ik \phi -t  k^{2} )$. In the limit
$\mathcal{I}_{0} \gg 1 $, this exact result simplifies to 
 \begin{eqnarray}
   \label{eq:DiffResult} 
   W( \mathcal{I}, \theta, \phi | t)  & = & 
   \frac{2 \exp (-2\mathcal{I}- 2\mathcal{I}_{0} + 4 \mathcal{I} \mathcal{I}_{0} )}
   {\pi^2\sqrt{\mathcal{I}\mathcal{I}_{0} \sin\theta \sin\theta_0}}
   \nonumber \\
  &\times &  \Theta\left ( \phi-\phi_0 - \chi t  \mathcal{I}   \cos \theta \Big| 
     \frac{\gamma  t}{4} + \frac{\cos[(\theta-\theta_0)/2]}{2 
       S  \mathcal{I}_{0} \sin\theta\sin\theta_0}  \right) \, .
 \end{eqnarray}

 The snapshots of the evolution of this Wigner function can be again
 appreciated in figure 4, with $\gamma = 2 \gamma_a = 2 \gamma_b = 0.5
 \chi$. While at the beginning one can only observe a very gentle
 difference with the non-dissipative case, this difference gets more
 visible as time goes by. The shrinking of the isolevels of the Wigner
 function for the dissipative evolution means that it gets ``smeared
 out'' over the phase space due to the dephasing. Note that the shape
 and the direction of the ellipsoids are not changed; only their size
 is different, indicating a lower degree of polarization.

We can also investigate the impact of dephasing  on squeezing. To this
end we need to calculate the corresponding quantities as in
equations~(\ref{eq:SyOfT}) and (\ref{eq:DeltaSTheta}). One finally gets
\begin{eqnarray} 
\label{eq:SyOfTDiff}
\fl
 \langle \op{S}_{y}(t) \rangle  = 
  \frac{ \mathcal{I}_{0}}{( 1+ \tau^{2} )^{2}}
  \exp \left ( - \frac{2\mathcal{I}_{0} \tau^{2}}
    {1 + \tau^{2}}  - \frac{\gamma t}{4}\right)  \\
 \fl 
 \Delta^{2} \op{S}_\perp ( \theta, t)  = 
\mathcal{I}_{0} [ 1 + (\mathcal{I}_{0}/2) \sin^2( \vartheta /2) ]  
-\sin^2\vartheta \frac{\mathcal{I}_{0}^2}{(1+4\tau^2)^3}\exp\left( 
-\frac{8 \mathcal{I}_{0}\tau^2}{1 + 4\tau^2} - \gamma t\right) &  \nonumber\\  
\fl
 \displaystyle 
  -  \sin(2\vartheta)\frac{2 \mathcal{I}_{0}\tau}{(1 + \tau^2)^3} 
\left( 1+ \frac{\mathcal{I}_{0}}{1 + \tau^2} \right)
\exp \left(
  -\frac{ 2 \mathcal{I}_{0}\tau^2}{1 + \tau^2} -\frac{\gamma t}{4}\right) \, . 
\end{eqnarray}
\begin{figure}
  \centering
  \includegraphics[height=6.5cm]{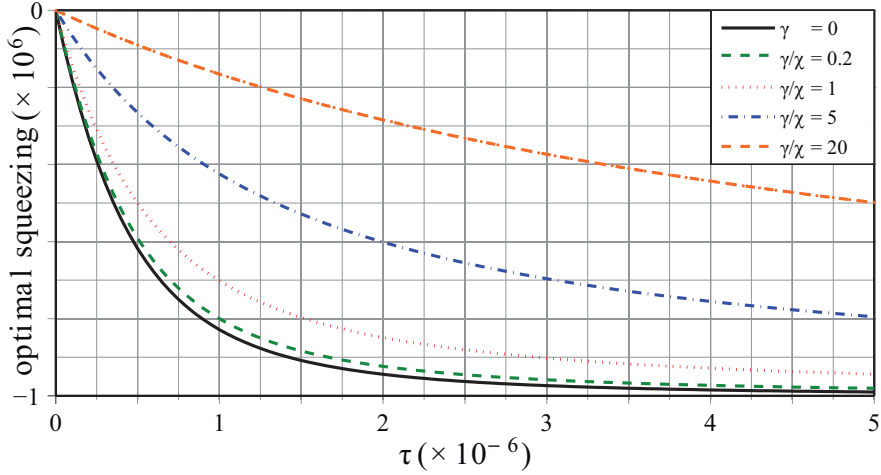}
  \caption{Optimal amount of squeezing $ \Delta^{2} \hat
    S_{\vartheta}^\mathrm{\,opt} (t) - |\langle \hat S_y (t)\rangle|
    $.  The values for $\gamma/\chi$ are 0 (black solid), 0.2 (green
    dashed), 1 (red dotted), 5 (blue dash-dotted) and 20 (orange
    dashed).}
\end{figure}
For short times $\tau \ll 1$ one can show that the optimal squeezing
angle is approximately given by
\begin{equation}
  \label{eq:ThetaOptDif}
  \vartheta_{\mathrm{sq}} \simeq
  \frac{1}{2}\mathrm{arccot}\left( \mathcal{I}_{0}\tau +
    \frac{\gamma}{4\chi} \right) \, .
\end{equation}
Note that in contradistinction with equation~(\ref{eq:OptSqzAngle}),
for a given time $\tau$ the optimal squeezing angle is closer to 0 in
the presence of dephasing. In a certain sense, dephasing makes the
isocontour ellipsoid rotate faster (yet also making it smaller).
Finally, the optimal squeezing amount turns out to be
\begin{equation}
  \label{eq:SqzOptDiff}
  \Delta^{2} \hat S_\perp^\mathrm{\,opt} (\vartheta, t)  - 
 |\langle  \hat S_y  (t)\rangle| 
  \simeq
2\mathcal{I}_{0}^2 \tau\left[ \mathcal{I}_{0} \tau + \frac{\gamma}{4\chi} -
  \sqrt{1 + \left( \mathcal{I}_{0} \tau + 
   \frac{ \gamma}{4\chi}\right)^2} \right] \, .
\end{equation}
In figure~4 we plot this optimal squeezing for several values of the
ratio $\gamma/\chi$. The degradation of this quantity with
$\gamma/\chi$ can be clearly observed.

\section{Concluding remarks}
\label{sec:conclusion}

In summary, we have presented a quasiclassical approximation to the
light propagation in a cross-Kerr medium. Even if the states considered
are bright and we neglect quantum correlations, we
still observe nonclassical effects such as entanglement or
squeezing. Interestingly, in the quasiclassical limit the correlations
remain in the system once induced, as opposed to the periodical
decorrelation observed in the exact evolution. We have also
constructed a model for dephasing processes in these media,
demonstrating that dissipation does visibly affect the degree of
polarization, but not so much its vectorial direction.


\ack

Financial support from the EU FP7 (Grant Q-ESSENCE), the Spanish DGI
(Grants FIS2008-04356 and FIS2011-26786), the UCM-BSCH program (Grant
GR-920992), and the Mexican CONACyT (Grant 106525) is acknowledged.

\begin{appendix}

\section{Two-mode Wigner function}
\label{ap:Wigner}

In this appendix a brief review of the Wigner distribution is given
for the problem at hand. For a single mode $a$, the Wigner function
for a state given by the density matrix $\op{\varrho}$ is
defined as
\begin{equation}
  \label{Ws}
  W (\alpha ) = 
  \Tr [ \op{\varrho}_{a} \, \op{w} (\alpha) ] \, , 
\end{equation}
where the kernel $\op{w} (\alpha)$ reads
\begin{equation}
  \op{w} (\alpha)  = \frac{1}{\pi^{2}} \int d\lambda \,
  \exp(\alpha \lambda^{\ast} - \alpha^{\ast} \lambda ) \, \op{D} (\alpha ) \, ,
\end{equation}
so it appears as the Fourier transform of the  displacement operator
$\op{D} (\alpha)$, with 
\begin{equation}
  \op{D} (\alpha ) =  \exp ( \alpha \op{a}^\dagger - 
  \alpha^\ast \op{a} ) \, .
\end{equation}
Note that the standard coherent states $| \alpha \rangle$ are generated
by  the action of $\op{D} (\alpha)$ on the vacuum, i.e.
\begin{equation}
  \label{eq:cs}
  | \alpha \rangle = \op{D} (\alpha)  | 0 \rangle \, .
\end{equation}
For a coherent state $| \alpha_{0} \rangle$, the Wigner function is
\begin{equation}
  \label{eq:Wcoh}
  W (\alpha) = \frac{2}{\pi} \exp ( - 2 | \alpha - \alpha_{0} | ) \, .
\end{equation}

In a more general context, the Wigner function can be
interpreted as the phase-space symbol of the density matrix
$\op{\varrho}$. This notion can be extended to any 
operator $\op{O}$ in such a way that its symbol is given by
\begin{equation}
  \label{Wsg}
  W_{\op{O}} (\alpha ) = 
  \Tr [ \op{O} \, \op{w} (\alpha) ] \, . 
\end{equation}
In particular, for the basic mode operator $\op{a}$ we have
\begin{equation}
  \label{eq:esa}
  W_{\op{a}} = \frac{\alpha}{\pi} \, .
 \end{equation}

In terms of $W( \alpha ) $, we can map any operator evolution into a
differential equation using the following rules~\cite{Gardiner2004}
\begin{eqnarray}
  \label{mapp}
  & \displaystyle
  \op{a} \op{\varrho}  \mapsto   \left ( 
    \alpha  +  \frac{1}{2}\frac{\partial}{\partial \alpha^{\ast}} 
  \right) \,  W (\alpha ) \, ,
  \qquad 
  \op{a}^{\dagger} \op{\varrho} \mapsto  
  \left ( 
    \alpha^\ast -  \frac{1}{2}\frac{\partial}{\partial \alpha} 
  \right ) \, W (\alpha ) \, , 
  &  \nonumber \\
  & & \\
  & \displaystyle
  \op{\varrho} \op{a}   \mapsto  
  \left ( 
    \alpha  -  \frac{1}{2}\frac{\partial}{\partial \alpha^{\ast}} 
  \right) W (\alpha )  \, , 
  \qquad
  \op{\varrho} \op{a}^{\dagger}  \mapsto  
  \left ( 
    \alpha^\ast +  \frac{1}{2}\frac{\partial}{\partial \alpha} 
  \right )  W (\alpha )  \, ,  \nonumber &
\end{eqnarray}
and after performing the decomposition (\ref{eq:abdec}), this reads
 \begin{eqnarray}
  \label{mappa}
  & \displaystyle
\delta  \op{a}  \,\op{\varrho}  \mapsto   
 \frac{1}{2}\frac{\partial}{\partial \alpha^{\ast}} 
\,  W \, ,
  \qquad \quad
\delta  \op{a}^{\dagger} \,
\op{\varrho} \mapsto  
 -  \frac{1}{2}\frac{\partial}{\partial \alpha} 
 \, W \, , 
  &  \nonumber \\
  & & \\
  & \displaystyle
  \op{\varrho} \,
\delta\op{a}   \mapsto  
 -  \frac{1}{2}\frac{\partial}{\partial \alpha^{\ast}} 
\, W \, , 
  \qquad 
  \op{\varrho} \,
\delta\op{a}^{\dagger}  \mapsto  
 \frac{1}{2}\frac{\partial}{\partial \alpha} 
\,W \, ,  \nonumber &
\end{eqnarray}
and analogous ones for the $b$ mode.

 The two-mode Wigner function is given by a direct generalization of 
equation (\ref{Ws}), namely 
\begin{equation}
  \label{W}
  W (\alpha, \beta) = 
  \Tr [ \op{\varrho} \, \op{w} (\alpha) \op{w} (\beta) ] \, . 
\end{equation}
The rest of the properties needed in the paper can be extended to this
two-mode case in a direct way.

\section{Purity of the reduced density matrix}
\label{ap:purity-reduc-dens}

For completeness, we give here some intermediate steps to obtain the
expression (\ref{Pex}) for the reduced purity $P_{a} (\tau)$, which is
defined as
\begin{equation}
  \label{eq:AppendixPurity}
  P_{a} (\tau)  =
  \frac{\pi}{8} \int_{-\pi}^{\pi} d\varphi_{a}
  \int_{0}^{\infty} d\mathcal{I}_{a}
  \left [ \int_{-\pi}^{\pi} d\varphi_{b} \int_{0}^{\infty} d\mathcal{I}_{b} 
    W(\mathcal{I}_{a},\varphi_{a} ;\mathcal{I}_{b} , \varphi_{b}|\tau)
  \right ]^{2} \, .
\end{equation}
Employing the form of the explicit form of the Wigner function
(\ref{Wt_0}) we have
\begin{eqnarray}
  \fl
  P_{a} (\tau) = \frac{2}{\pi^3} \int_{-\pi}^\pi   d\varphi_a \,  d\varphi_1 \,
  d\varphi_2  \int_0^{\infty}  d\mathcal{I}_{a} \,  d\mathcal{I}_{1} \, 
  d\mathcal{I}_{2} \,
  \exp ( -4\mathcal{I}_{a} - 4\mathcal{I}_{0a} - 2\mathcal{I}_1 -
  2\mathcal{I}_2 - 4\mathcal{I}_{0b} ) \nonumber  \\ 
  \fl
  \times  \exp\left[ 2\sqrt{\mathcal{I}_a \mathcal{I}_{0a} }\left( 
      e^{i\varphi_a +2 i\mathcal{I}_1 \tau} + e^{i\varphi_a +2 i\mathcal{I}_2 \tau} +
      e^{-i\varphi_a  - 2 i\mathcal{I}_1 \tau} + e^{-i\varphi_a - 2 i\mathcal{I}_2 \tau}
    \right) \right]  \nonumber \\
  \fl 
  \times \exp (4\sqrt{\mathcal{I}_1 \mathcal{I}_{0b} } \cos \varphi_1
  + 4\sqrt{\mathcal{I}_2 \mathcal{I}_{0b}}  \cos
  \varphi_2) \, .
  \label{eq:FullIntegrals}
\end{eqnarray}
In the second line, we can expand the exponential in power series in
$e^{i\varphi_a}$ and $e^{-i\varphi_a}$, considering the rest of
variables as fixed coefficients. Then, the integration over
$\varphi_a$ can be explicitly carried out, with the result
\begin{equation}
  2 \pi \sum_{k=0}^\infty \frac{(4\mathcal{I}_a)^k
    (2\mathcal{I}_{0a})^k}{(k!)^2} \left \{  1+ 
    \cos\left [ 2( \mathcal{I}_1 - \mathcal{I}_2) \tau \right ]  \right \}
  \, .
\end{equation}
Together with the term $e^{-4\mathcal{I}_a}$, this can be immediately
integrated over $d\mathcal{I}_a$, yielding
\begin{equation}
  \label{eq:expofcos}
  \frac{\pi}{4} \exp\left\{ 2\mathcal{I}_{0a} \left[ 1+ \cos\left(
        2(\mathcal{I }_1 - \mathcal{I}_2)
        \tau\right) \right]\right\} ,
\end{equation}
which replaces the second line in equation~(\ref{eq:FullIntegrals}).
The integrations over $\varphi_j$ transform the last line of
(\ref{eq:FullIntegrals}) into
\begin{equation}
  (2 \pi)^2  I_0 (4\sqrt{\mathcal{I}_1 \mathcal{I}_{0b} }) \,  I_0
  (4\sqrt{\mathcal{I}_2 \mathcal{I}_{0b} }) \,.
\end{equation}
Finally,  to carry out the integrations over $\mathcal{I}_1$ and
$\mathcal{I}_2$, we expand the Bessel functions in power series, namely
\begin{equation}
I_0 (4\sqrt{\mathcal{I}_j \mathcal{I}_{0b} }) = \sum_{k=0}^\infty
\frac{(2\mathcal{I }_j)^k
  (2\mathcal{I}_{0b})^k}{(k!)^2}\, ,
\end{equation}
as well as  the exponential in (\ref{eq:expofcos})
\begin{equation}
\exp\left[  2\mathcal{I}_{0a} \cos\left(2 (\mathcal{I}_1 - \mathcal{I}_2)
      \tau\right) \right] = \sum_{n = - \infty}^{\infty} I_ n(2\mathcal{I}_{0a} )
  \, \exp\left [ 2i (\mathcal{I}_1 - \mathcal{I}_2)  \tau \right ] \, .
\end{equation}
All this enables a direct  integration
over $\mathcal{I}_1$ and $\mathcal{I}_2$, getting
\begin{eqnarray}
  \label{eq:finalPurity} 
  P(\tau) & = & 4\exp ( -4\mathcal{I}_{0b}- 2\mathcal{I}_{0a} ) \int_0^\infty
  d\mathcal{I}_1 d\mathcal{I }_2 \,
  \exp( -2\mathcal{I}_1 - 2\mathcal{I}_2 ) \nonumber \\
 & \times & \sum_{k,m=0}^\infty
  \frac{(2\mathcal{I}_1)^k (2\mathcal{I}_2)^m
    (2\mathcal{I}_{0b})^{k+m}}{(k!)^2(m!)^2}  
\sum_{n=- \infty}^{\infty} I_n (2\mathcal{I}_{0a}) \, e^{2i(\mathcal{I}_1 -
    \mathcal{I}_2) \tau } \, .
\end{eqnarray}
From this expression, the result (\ref{Pex}) for the purity follows
straightforwardly. 

\end{appendix}

\newpage


\begin{thebibliography}{100}
\expandafter\ifx\csname url\endcsname\relax
  \def\url#1{{\tt #1}}\fi
\expandafter\ifx\csname urlprefix\endcsname\relax\def\urlprefix{URL }\fi
\providecommand{\eprint}[2][]{\url{#2}}

\bibitem{Braginsky:1968fk}
Braginsky V~B 1968 {\em Sov. Phys. JETP\/} {\bf 26} 831--834

\bibitem{Unruh:1979uq}
Unruh W~G 1979 {\em Phys. Rev. D\/} {\bf 19} 2888--2896

\bibitem{Milburn:1983kx}
Milburn G~J and Walls D~F 1983 {\em Phys. Rev. A\/} {\bf 28} 2065--2070

\bibitem{Imoto:1985vn}
Imoto N, Haus H~A and Yamamoto Y 1985 {\em Phys. Rev. A\/} {\bf 32} 2287--2292

\bibitem{Alsing:1988ys}
Alsing P, Milburn G~J and Walls D~F 1988 {\em Phys. Rev. A\/} {\bf 37}
  2970--2978

\bibitem{Grangier:1998zr}
Grangier P, Levenson J~A and Poizat J~P 1998 {\em Nature\/} {\bf 396} 537--542

\bibitem{Sanders:1989ly}
Sanders B~C and Milburn G~J 1989 {\em Phys. Rev. A\/} {\bf 39} 694--702

\bibitem{Konig:2002uq}
K\"{o}nig F, B\"{u}chler B, Rechtenwald T, Leuchs G and Sizmann A 2002 {\em
  Phys. Rev. A\/} {\bf 66} 043810

\bibitem{Xiao:08}
Xiao Y~F, \"{O}zdemir S~K, Gaddam V, Dong C~H, Imoto N and Yang L 2008 {\em
  Opt. Express\/} {\bf 16} 21462--21475

\bibitem{Milburn:1986ve}
Milburn G~J and Holmes C~A 1986 {\em Phys. Rev. Lett.\/} {\bf 56} 2237--2240

\bibitem{Yurke:1986qf}
Yurke B and Stoler D 1986 {\em Phys. Rev. Lett.\/} {\bf 57} 13--16

\bibitem{Tombesi:1987bh}
Tombesi P~and~Mecozzi A 1987 {\em J. Opt. Soc. Am. B\/}  1700--1709

\bibitem{Gantsog:1991cr}
Gantsog T and Tana{\'s} R 1991 {\em J. Mod. Opt.\/} {\bf 38} 1537--1558

\bibitem{Wilson-Gordon:1991bs}
Wilson-Gordon A~D, Bu\v{z}ek V and Knight P~L 1991 {\em Phys. Rev. A\/} {\bf
  44} 7647--7656

\bibitem{Tara:1993nx}
Tara K, Agarwal G~S and Chaturvedi S 1993 {\em Phys. Rev. A\/} {\bf 47}
  5024--5029

\bibitem{Luis:1995kl}
Luis A, S{\'a}nchez-Soto L~L and Tana{\'s} R 1995 {\em Phys. Rev. A\/} {\bf 51}
  1634--1643

\bibitem{Szabo:1996oq}
Szabo S, Adam P, Janszky J and Domokos P 1996 {\em Phys. Rev. A\/} {\bf 53}
  2698--2710

\bibitem{Chumakov:1999tg}
Chumakov S~M, Frank A and Wolf K~B 1999 {\em Phys. Rev. A\/} {\bf 60}
  1817--1823

\bibitem{Korolkova:2001ij}
Korolkova N, Loudon R, Gardavsky G, Hamilton M~W and Leuchs G 2001 {\em J. Mod.
  Opt.\/} {\bf 48} 1339--1355

\bibitem{Vitali:2000fv}
Vitali D, Fortunato M and Tombesi P 2000 {\em Phys. Rev. Lett.\/} {\bf 85}
  445--448

\bibitem{Jian:2009dz}
Jian Z, Ming Y, Yan L and Zhuo-Liang C 2009 {\em Chinese Phys. Lett.\/} {\bf
  26} 100301

\bibitem{Zhu:2011qa}
Zhu M, Yin X and Yuan G 2011 {\em Opt. Commun.\/} {\bf 284} 3483--3487

\bibitem{Turchette:1995fu}
Turchette Q~A, Hood C~J, Lange W, Mabuchi H and Kimble H~J 1995 {\em Phys. Rev.
  Lett.\/} {\bf 75} 4710--4713

\bibitem{Semiao:2005fu}
Semi\~{a}o F~L and Vidiella-Barranco A 2005 {\em Phys. Rev. A\/} {\bf 72}
  064305

\bibitem{Matsuda:2007ye}
Matsuda N, Mitsumori Y, Kosaka H, Edamatsu K and Shimizu R 2007 {\em Appl.
  Phys. Lett.\/} {\bf 91} 171119

\bibitem{Azuma:2008qo}
Azuma H 2008 {\em J. Phys. D\/} {\bf 41} 025102

\bibitem{You:2011pi}
You H and Franson J 2011 {\em Quantum Inf. Process.\/}  1--25

\bibitem{Xia:2011ff}
Xia Y, Song J, Lu P~M and Song H~S 2011 {\em J. Phys. B\/} {\bf 44} 025503

\bibitem{Kok:2002tw}
Kok P, Lee H and Dowling J~P 2002 {\em Phys. Rev. A\/} {\bf 66} 063814

\bibitem{Munro:2005il}
Munro W~J, Nemoto K, Beausoleil R~G and Spiller T~P 2005 {\em Phys. Rev. A\/}
  {\bf 71} 033819

\bibitem{Greentree:2009jl}
Greentree A~D, Beausoleil R~G, Hollenberg L~C~L, Munro W~J, Nemoto K, Prawer S
  and Spiller T~P 2009 {\em New J. Phys.\/} {\bf 11} 093005

\bibitem{Schmidt:1996oa}
Schmidt H and Imamo\u{g}lu A 1996 {\em Opt. Lett.\/} {\bf 21} 1936--1938

\bibitem{Imamoglu:1997sh}
Imamo\u{g}lu A, Schmidt H, Woods G and Deutsch M 1997 {\em Phys. Rev. Lett.\/}
  {\bf 79} 1467--1470

\bibitem{Werner:1999xd}
Werner M~J and Imamo\u{g}lu A 1999 {\em Phys. Rev. A\/} {\bf 61} 011801

\bibitem{Dey:2007wb}
Dey T~N and Agarwal G~S 2007 {\em Phys. Rev. A\/} {\bf 76} 015802

\bibitem{Hau:1999xz}
Hau L, Harris S, Dutton Z and Behroozi C 1999 {\em Nature\/} {\bf 397} 594--598

\bibitem{Kang:2003le}
Kang H and Zhu Y 2003 {\em Phys. Rev. Lett.\/} {\bf 91} 093601

\bibitem{Castellanos:2007kx}
Castellanos-Beltran M~A and Lehnert K~W 2007 {\em Appl. Phys. Lett.\/} {\bf 91}
  083509

\bibitem{Mallet:2009uq}
Mallet F, Ong F~R, Palacios-Laloy A, Nguyen F, Bertet P, Vion D and Esteve D
  2009 {\em Nat. Phys.\/} {\bf 5} 791--795

\bibitem{Bergeal:465fk}
Bergeal N, Schackert F, Metcalfe M, Vijay R, Manucharyan V~E, Frunzio L, Prober
  D~E, Schoelkopf R~J, Girvin S~M and Devoret M~H 2010 {\em Nature\/} {\bf 465}
  64--68

\bibitem{Bermel:2007jb}
Bermel P, Rodriguez A, Joannopoulos J~D and Soljaƒ{\c c}iƒ{\'a} M 2007 {\em
  Phys. Rev. Lett.\/} {\bf 99} 053601

\bibitem{Mohapatra:2008xi}
Mohapatra A~K, Bason M~G, Butscher B, Weatherill K~J and Adams C~S 2008 {\em
  Nature Phys.\/} {\bf 4} 890--894

\bibitem{Brandao:2008mq}
Brand\~{a}o F~G~S~L, Hartmann M~J and Plenio M~B 2008 {\em New J. Phys.\/} {\bf
  10} 043010

\bibitem{Babourina-Brooks:2008wm}
Babourina-Brooks E, Doherty A and Milburn G~J 2008 {\em New J. Phys.\/} {\bf
  10} 105020

\bibitem{Ritze:1979mb}
Ritze H~H and Bandilla A 1979 {\em Opt. Commun.\/} {\bf 29} 126--130

\bibitem{Heiman:1976cq}
Heiman D, Hellwarth R~W, Levenson M~D and Martin G 1976 {\em Phys. Rev.
  Lett.\/} {\bf 36} 189--192

\bibitem{Yuen:1979rq}
Yuen H~P and Shapiro J~H 1979 {\em Opt. Lett.\/} {\bf 4} 334--336

\bibitem{Levenson:1985fc}
Levenson M~D, Shelby R~M and Perlmutter S~H 1985 {\em Opt. Lett.\/} {\bf 10}
  514--516

\bibitem{Levenson:1985ss}
Levenson M~D, Shelby R~M, Aspect A, Reid M and Walls D~F 1985 {\em Phys. Rev.
  A\/} {\bf 32} 1550--1562

\bibitem{Kitagawa:1986qc}
Kitagawa M and Yamamoto Y 1986 {\em Phys. Rev. A\/} {\bf 34} 3974--3988

\bibitem{Joneckis:1993bd}
Joneckis L~G and Shapiro J~H 1993 {\em J. Opt. Soc. Am. B\/} {\bf 10}
  1102--1120

\bibitem{Sundar:1996ud}
Sundar K 1996 {\em Phys. Rev. A\/} {\bf 53} 1096--1111

\bibitem{Schmitt:1998hs}
Schmitt S, Ficker J, Wolff M, K{\"o}nig F, Sizmann A and Leuchs G 1998 {\em
  Phys. Rev. Lett.\/} {\bf 81} 2446--2449

\bibitem{Boyd:1999kc}
Boyd R~W 1999 {\em J. Mod. Opt.\/} {\bf 46} 367--378

\bibitem{Shelby:1985sw}
Shelby R~M, Levenson M~D and Bayer P~W 1985 {\em Phys. Rev. B\/} {\bf 31}
  5244--5252

\bibitem{Elser:2006fh}
Elser D, Andersen U~L, Korn A, Gl{\"o}ckl O, Lorenz S, Marquardt C and Leuchs G
  2006 {\em Phys. Rev. Lett.\/} {\bf 97} 133901

\bibitem{Agrawal:2007lp}
Agrawal G~P 2007 {\em Nonlinear Fiber Optics\/} 4th ed (New York: Academic)

\bibitem{Lee:1995wa}
Lee H~W 1995 {\em Phys. Rep.\/} {\bf 259} 147--211

\bibitem{Schroek:1996ta}
Schroek F~E 1996 {\em Quantum Mechanics on Phase Space\/} (Dordrecht: Kluwer)

\bibitem{Schleich:2001bu}
Schleich W~P 2001 {\em Quantum Optics in Phase Space\/} (Berlin: Wiley-VCH)

\bibitem{Heller:1976yj}
Heller E~J 1976 {\em J. Chem. Phys.\/} {\bf 65} 1289--1298

\bibitem{Lee:1980bf}
Lee H~W and Scully M~O 1980 {\em J. Chem. Phys.\/} {\bf 73} 2238--2242

\bibitem{Balzer:2003wq}
Balzer B, Dilthey S, Stock G and Thos M 2003 {\em J. Chem. Phys.\/} {\bf 119}
  5795--5804

\bibitem{Hillery:1984ez}
Hillery M, O'Connell R~F, Scully M~O and Wigner E~P 1984 {\em Phys. Rep.\/}
  {\bf 106} 121--167

\bibitem{Drobny:1997tp}
Drobn{\'y} G, Bandilla A and Jex I 1997 {\em Phys. Rev. A\/} {\bf 55} 78--93

\bibitem{Banaszek:1997km}
Banaszek K and Knight P~L 1997 {\em Phys. Rev. A\/} {\bf 55} 2368--2375

\bibitem{Bandilla:2000qw}
Bandilla A, Drobn{\'y} G and Jex I 2000 {\em J. Opt. B\/} {\bf 2} 265--270

\bibitem{Stobinska:2008vz}
Stobi\'{n}ska M, Milburn G~J and W\'{o}dkiewicz K 2008 {\em Phys. Rev. A\/}
  {\bf 78} 013810

\bibitem{Corney:2008zr}
Corney J~F, Heersink J, Dong R, Josse V, Drummond P~D, Leuchs G and Andersen
  U~L 2008 {\em Phys. Rev. A\/} {\bf 78} 023831

\bibitem{Sizmann:1999if}
Sizmann A and Leuchs G 1999 The optical {K}err effect and quantum optics in
  fibers {\em Progress in Optics\/} vol~39 ed Wolf E (Amsterdam: Elsevier) pp
  373--469

\bibitem{Duan:2000ly}
Duan L~M, Giedke G, Cirac J~I and Zoller P 2000 {\em Phys. Rev. Lett.\/} {\bf
  84} 2722--2725

\bibitem{Simon:2000ve}
Simon R 2000 {\em Phys. Rev. Lett.\/} {\bf 84} 2726--2729

\bibitem{Weiss:1999vn}
Weiss U 1999 {\em Quantum Dissipative Systems\/} (Singapore: World Scientific)

\bibitem{Scully:2001lp}
Scully M~O and Zubairy M~S 2001 {\em Quantum Optics\/} (Cambridge: Cambridge
  University Press)

\bibitem{Goldstein:1980fk}
Goldstein H 1980 {\em Classical Mechanics\/} (New York: Addison-Wesley)

\bibitem{Heersink:2003oz}
Heersink J, Gaber T, Lorenz S, Gl{\"o}ckl O, Korolkova N and Leuchs G 2003 {\em
  Phys. Rev. A\/} {\bf 68} 013815

\bibitem{Heersink:2005ul}
Heersink J, Josse V, Leuchs G and Andersen U~L 2005 {\em Opt. Lett.\/} {\bf 30}
  1192--1194

\bibitem{Braunstein:2005fk}
Braunstein S~L and van Loock P 2005 {\em Rev. Mod. Phys.\/} {\bf 77} 513--577

\bibitem{Paris:2005uq}
Ferraro A, Olivares S and Paris M~G~A 2005 {\em Gaussian States in Quantum
  Information\/} (Naples: Bibliopolis)

\bibitem{Wanga:2007ys}
Wanga X~B, Hiroshima T, Tomita A and Hayashi M 2007 {\em Phys. Rep.\/} {\bf
  448} 1---111

\bibitem{Adesso:2007zr}
Adesso G and Illuminati F 2007 {\em J. Phys. A\/} {\bf 40} 7821--7880

\bibitem{Weedbrook:2012kx}
Weedbrook C, Pirandola S, Garc{\'\i}a-Patr{\'o}n R, Cerf N~J, Ralph T~C,
  Shapiro J~H and Lloyd S 2012 {\em Rev. Mod. Phys.\/} {\bf 84} 621--669

\bibitem{Shchukin:2005bh}
Shchukin E and Vogel W 2005 {\em Phys. Rev. Lett.\/} {\bf 95} 230502

\bibitem{Agarwal:2005dq}
Agarwal G~S and Biswas A 2005 {\em New J. Phys.\/} {\bf 7} 211

\bibitem{Hillery:2006nx}
Hillery M and Zubairy M~S 2006 {\em Phys. Rev. Lett.\/} {\bf 96} 050503

\bibitem{Gomes:2009qf}
Gomes R~M, Salles A, Toscano F, Souto~Ribeiro P~H and Walborn S~P 2009 {\em
  Proc. Natl. Acad. Sci. USA\/} {\bf 106} 21517--21520

\bibitem{Mazzola:2010ly}
Mazzola L, Bellomo B, Lo~Franco R and Compagno G 2010 {\em Phys. Rev. A\/} {\bf
  81} 052116

\bibitem{De-Pasquale:2010ys}
De~Pasquale A, Facchi P, Parisi G, Pascazio S and Scardicchio A 2010 {\em Phys.
  Rev. A\/} {\bf 81} 052324

\bibitem{De-Pasquale:2012zr}
Pasquale A~D, Facchi P, Giovannetti V, Parisi G, Pascazio S and Scardicchio A
  2012 {\em J. Phys. A\/} {\bf 45} 015308

\bibitem{Abramowitz:1984}
Abramowitz M and Stegun I~A (eds) 1984 {\em Handbook of Mathematical
  Functions\/} (New York: Dover)

\bibitem{Luis:2000ys}
Luis A and S{\'a}nchez-Soto L~L 2000 Quantum phase difference, phase
  measurements and {S}tokes operators {\em Progress in Optics\/} vol~41 ed Wolf
  E (Amsterdam: Elsevier) pp 421--481

\bibitem{Sehat:2005wd}
Sehat A, S{\"{o}}derholm J, Bj{\"{o}}rk G, Espinoza P, Klimov A~B and
  S{\'a}nchez-Soto L~L 2005 {\em Phys. Rev. A\/} {\bf 71} 033818

\bibitem{Marquardt:2007bh}
Marquardt C, Heersink J, Dong R, Chekhova M~V, Klimov A~B, S{\'a}nchez-Soto
  L~L, Andersen U~L and Leuchs G 2007 {\em Phys. Rev. Lett.\/} {\bf 99} 220401

\bibitem{Bjork:2010rt}
Bj{\"{o}}rk G, S{\"{o}}derholm J, S{\'a}nchez-Soto L~L, Klimov A~B, Ghiu I,
  Marian P and Marian T~A 2010 {\em Opt. Commun.\/} {\bf 283} 4440--4447

\bibitem{Muller:2012mk}
M{\"u}ller C~R, Stoklasa B, Peuntinger C, Gabriel C, \v{R}eh\'a\v{c}ek J,
  Hradil Z, Klimov A~B, Leuchs G, Marquardt C and S{\'{a}}nchez-Soto L~L 2012
  {\em New J. Phys.\/} {\bf 14} 085002

\bibitem{Chirkin:1993dz}
Chirkin A~S, Orlov A~A and Parashchuk D~Y 1993 {\em Quantum Electron.\/} {\bf
  23} 870--874

\bibitem{Korolkova:2002fu}
Korolkova N, Leuchs G, Loudon R, Ralph T~C and Silberhorn C 2002 {\em Phys.
  Rev. A\/} {\bf 65} 052306

\bibitem{Luis:2006ye}
Luis A and Korolkova N 2006 {\em Phys. Rev. A\/} {\bf 74} 043817

\bibitem{Mahler:2010fk}
Mahler D, Joanis P, Vilim R and de~Guise H 2010 {\em New J. Phys.\/} {\bf 12}
  033037

\bibitem{Ma:2011ve}
Ma J, Wang X, Sun C~P and Nori F 2011 {\em Phys. Rep.\/} {\bf 509} 89--165

\bibitem{Klimov:2006fd}
Klimov A~B, Delgado J and S{\'{a}}nchez-Soto L~L 2006 {\em Opt. Commun.\/} {\bf
  258} 210--218

\bibitem{Zurek:2003fk}
Zurek W~H 2003 {\em Rev. Mod. Phys.\/} {\bf 75} 715--775

\bibitem{Shao:1996uq}
Shao J, Ge M~L and Cheng H 1996 {\em Phys. Rev. E\/} {\bf 53} 1243--1245

\bibitem{Mozyrsky:1998kx}
Mozyrsky D and Privman V 1998 {\em J. Stat. Phys.\/} {\bf 91} 787--799

\bibitem{Gardiner2004}
Gardiner C~W and Zoller P 2004 {\em Quantum Noise\/} 2nd ed (Berlin: Springer)

\bibitem{Breuer2002}
Breuer H~P and Petruccione F 2002 {\em The Theory of Open Quantum Systems\/}
  (Oxford: Oxford University Press)

\bibitem{Klimov:2008ys}
Klimov A~B, Romero J~L, S{\'a}nchez-Soto L~L, Messina A and Napoli A 2008 {\em
  Phys. Rev. A\/} {\bf 77} 033853

\end{thebibliography}

\providecommand{\newblock}{}

\end{document}